\RequirePackage{fix-cm}
 \documentclass[final,authoryear,5p,times,twocolumn]{elsarticle}
\usepackage{graphicx}
\usepackage{color}
 \journal{Icarus}
\begin{document}
\begin{frontmatter}



\title{Amplification of dust loading in Martian dust devils by self-shadowing}


\author{M. Kuepper}
\author{G. Wurm}
\ead{gerhard.wurm@uni-due.de}

\address{Faculty of Physics, University of Duisburg-Essen, Lotharstr. 1, 47057 Duisburg, Germany}

\begin{abstract}
Insolation of the Martian soil leads to a sub-surface overpressure due
to thermal creep gas flow. This could support particle entrainment into the atmosphere.
Short time shadowing e.g. by the traverse of a larger dust devil would enhance this effect. 
We find in microgravity experiments that mass ejection rates are increased 
by a factor of 10 for several seconds if a light source of 12.6\,kW/m$^{2}$ is turned off.
Scaled to Mars this implies that self-shadowing of a partially opaque dust devil might
lead to a strongly amplified flux of lifted material. We therefore suggest that 
self-shadowing might be a mechanism on Mars to increase the total dust loading 
of a dust devil and keep it self-sustained.\\


%\keywords{Mars, surface \and Mars, Atmoshere \and Atmospheres, dynamics \and Dust Devils}
\end{abstract}

\end{frontmatter}

\section{Introduction}
\label{intro}
It is still a challenge to explain dust entrainment within Martian dust devils. A current overview of lifting mechanisms can be found in \cite{Neakrase2016}. It includes gas drag \cite{Greeley1980}, pressure differences associated to a dust devils passage \cite{Balme2006} and thermal creep induced overpressures \cite{deBeule2014, Kuepper2015}. However, it is unclear if the necessary conditions (wind speed, pressure difference) are always met in the low pressure atmosphere on Mars.

This paper is not treating the initial formation of a dust devil in spite of all possible problems 
but assumes it to be existent. There is no doubt that dust devils can form as they have been observed
frequently by now in different locations e.g. by rovers or satellite images \cite{Greeley2006a,Reiss2014b}.

While small dust devils might not be totally opaque, observations of larger devils show that they can be optically thick, casting shadows as they move
\cite{Greeley2006a,Reiss2014b}.  Depending on the size of a dust devil and its speed it takes several seconds to
cross a spot of Martian soil along the trajectory. This implies that this spot, which is illuminated
before the dust devil's arrival is shadowed for several seconds during the devil's passage.
This change in illumination might have a severe impact on the dust flux lifted from the ground
which is the focus of this work.

\section{Illumination and lifting}

By insolation the dust is heated. This heat can partly be reradiated from the surface as thermal radiation. The absorption length of infrared radiation is normally much shorter than for visible wavelength~--- therefore only the surface can cool efficiently, but heat is deposited along several layers at the top of the dustbed. Inside the dustbed thermal conduction and radiation between the single grains have to be considered. This results in a relatively flat temperature profile at the top and a decline to ambient temperature deeper inside the dust bed. If the illumination is switched off, the thermal radiation will quickly cool the surface, leading to a temperature profile with a maximum inside the dust bed.

In rarefied gases temperature gradients have peculiar consequences. The Knudsen effect \cite{Knudsen1909} is important here, as it states that in equilibrium the pressures $p$ in two chambers communicating through a capillary smaller than the mean free path of the gas is determined by their temperatures $T$
\begin{equation}
\frac{p_1}{p_2}=\sqrt{\frac{T_1}{T_2}}\,.
\end{equation}

If not in equilibrium, the associated mass flow rate of the gas $\dot{M}$ can be described for an indefinite capillary with a temperature gradient as \citep{Muntz2002} 
\begin{equation}
\dot{M} = p_{\mathrm{avg}} \frac{A}{\sqrt{2 \frac{k_B}{m} T_{\mathrm{avg}}}} \left(\frac{L_r}{T_{\mathrm{avg}}} \frac{\Delta T}{L_x} Q_T - \frac{L_r}{p_{\mathrm{avg}}}\frac{\Delta p}{L_x} Q_p \right)\,,
\end{equation}
where $p_{\mathrm{avg}}$ is the average pressure in the capillary, $T_{\mathrm{avg}}$ the average temperature, $A$ is the cross section of the capillary, $L_r$ its radius, $L_x$ its length, $k_B$ is the Boltzmann constant, $m$ the molecular mass of the gas, $\Delta T$ is the temperature difference across the capillary, $\Delta p$ the pressure difference, $Q_T$ and $Q_p$ are the coefficients of thermal and pressure driven flow and depend on the Knudsen number (see \cite{Sone1990}). 
%This applies lifting forces to the grains of a dust bed.

\begin{figure}
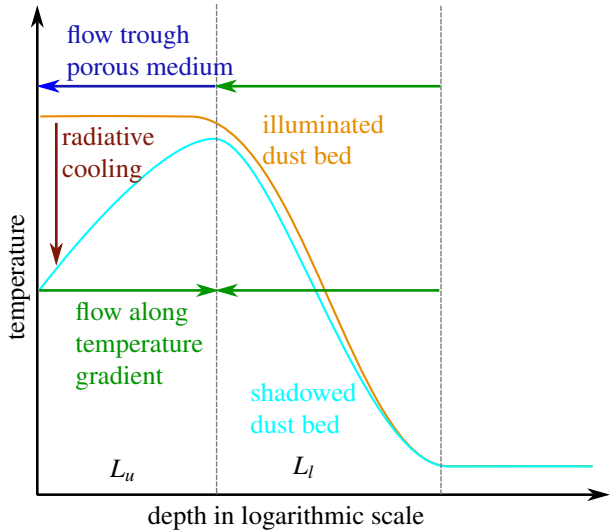
\caption{Sketch of the influence of a shadowed region. A detailed calculation is done in \cite{Kocifaj2011}. For the illuminated dust bed the temperature gradient induced thermal creep pumps gas from the deeper layers upwards. The flat temperature profile at the top leads to an overpressure inside the dust bed, as there is only pressure driven flow through the pores here. If the illumination is switched off, the surface cools by thermal radiation. This reversed temperature gradient enhances the sub-surface pressure.}
\label{fig:sketch}
\end{figure}

The details will depend in a complex way on particle optical properties, dust bed morphology, 
time evolution of the ejections themselves, heat transfer and so on. We do not aim to construct a detailed model at this point but just quantify the ratio of ejected particle rates with
and without illumination in measurements. Therefore, experiments have been carried out in microgravity at the drop tower in Bremen, where ejected particles can be observed without gravitational bias. 
Additional experiments using a centrifuge in parabolic flights provided data for low gravity of 0.3g, close to Martian gravity. 

\section{Drop tower experiments}

\begin{figure}
\includegraphics[width=0.5\textwidth]{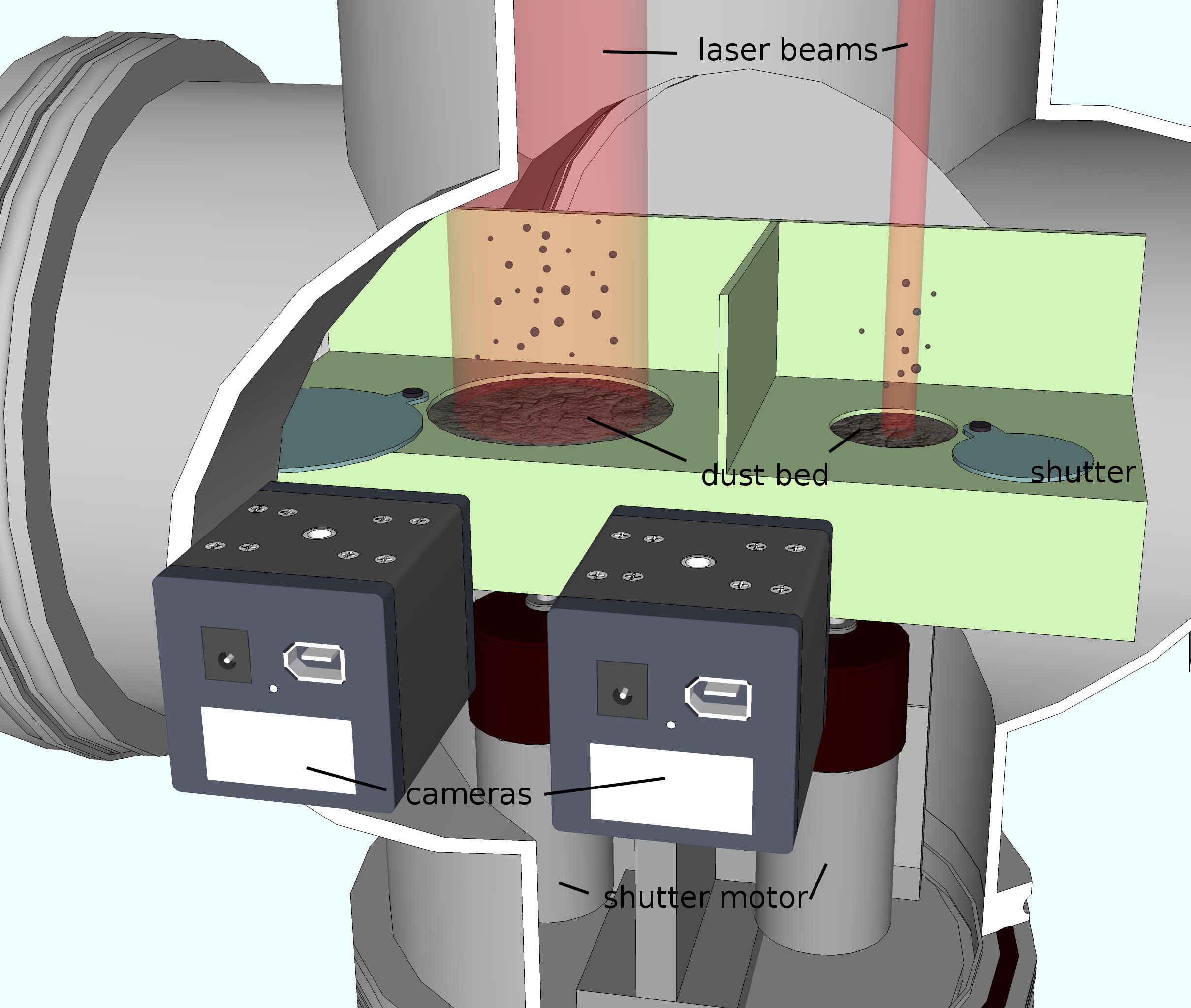}
\caption{Sketch of the experiment. Two different dust beds are placed in a vacuum chamber and illuminated with lasers from above. During catapult launch, the dust is protected by a shutter.
\label{sketchi}}
\end{figure}

The principle of the experiment is straight forward (fig. \ref{sketchi}). The microgravity time for an experiment is about 9\,s. 

A dust bed is sealed during launch of the experiment capsule by a lid. The dust bed is placed within a vacuum chamber which is evacuated to 2\,mbar pressure. Once microgravity is reached the cover is removed. The dust is then exposed to illumination. Illumination is provided by a diode laser. The laser is coupled into a glass fiber. The exiting light beam is shaped to provide an essentially parallel beam with a certain spot diameter on the dust bed surface.  
This illumination leads to particle ejections from the dust bed. Particles are observed by a camera from the side. The frame rate varied for the experiments considered between the two values of 1000\,fps and 500\,fps. The spatial resolution of the optical system is 68 $\rm \mu m$/pixel. Illumination for the observations is provided by two means. First, scattered laser light provides imaging for particles ejected. In addition an illuminated background provides dark images of particles not within the laser light. During the course of an individual experiment the laser is turned off. Particles are now still visible in front of the illuminated background.
Two slightly different configurations were used, one with a large infrared laserspot (955\,nm wavelength, 3.4\,cm spot diameter, 7\,cm dust bed diameter, 2\,cm dust bed depth, radiant flux at the surface of 12.7\,kW/m$^2$) and one with a smaler red laserspot (655\,nm wavelength, 5.5\,mm spot diameter, 3\,cm dust bed diameter, 2\,cm dust bed depth, radiant flux at the surface of 12.6\,kW/m$^2$).

{The light sources were selected due to availability and limited number of microgravity experiments. Also, the high power available in IR allowed a larger spot diameter. It should be noted that this illumination differs from the solar spectrum. Ground based experiments of dust lifting with green and blue light do not suggest large differences in the relevant wavelength range but those are unpublished (de Beule, personal communication). Therefore, slight changes might have to be expected.}

The dust sample was a JSC-1A Mars simulant (palagonite) which was heated for 1\,h at 400$^\circ$K to remove residual volatile components. The samples had a broad size distribution below 1mm. The average particle size was 306 $\rm \mu m$.

\section{Data Reduction}

\begin{figure}
\includegraphics[width=0.5\textwidth]{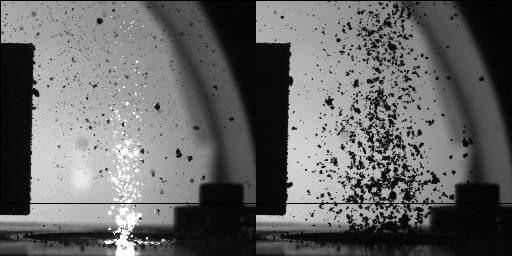}
\caption{Sample images of particle ejections with illumination (left) and after the light source is turned off (right) with the small laser spot. The region of interest, where particles are counted, is the whole area above the black line. Each image has a size of 3.4 cm.}
\label{sample}
\end{figure}

The light flux used induces particle ejections during illumination and an increased particle ejection
once the light source is turned off. Sample images from the video sequence are shown in fig. \ref{sample}

For the different experiments we used the following algorithm to deduce a ration between
the ejected mass flux with and without illumination.
From the video images with open lid at zero gravity, a median image was calculated. This background was subtracted from every image. The resulting images were binarized by setting a fixed brightness threshold. This outlines the particles illuminated by the laser. A second set of binarized images was generated in the same way to outline particles not illuminated by the laser by inverting the images first. Bright and dark particles were combined. Further image processing then provides the number of particles $n$ and the average size $A_P$ of individual particles. These are determined in a region of interest, which is always placed in the same way above the dust bed surface. This is also marked in fig. \ref{sample}. These data were used to estimate the ejected volume of particles by
\begin{equation}
V=\frac{4}{3}n \pi A_p^2\sqrt{\frac{A_p}{\pi}}\,.
\end{equation}

We neglect the fact that particles can shadow each other in dense regions. 
Due to different thresholds chosen for the two kinds of illumination the volume is discontinuous at the moment the light is switched off. Particles appeared larger while beeing illuminated. To account for this the "light-off" part was scaled to the "light-on" part right at the moment of switch-off, where 
no effects are visible yet. We note that
we are only interested in ratios between volumes later on and the exact setting of the threshold
is of minor importance.

\begin{figure}
  \includegraphics[width=0.5\textwidth]{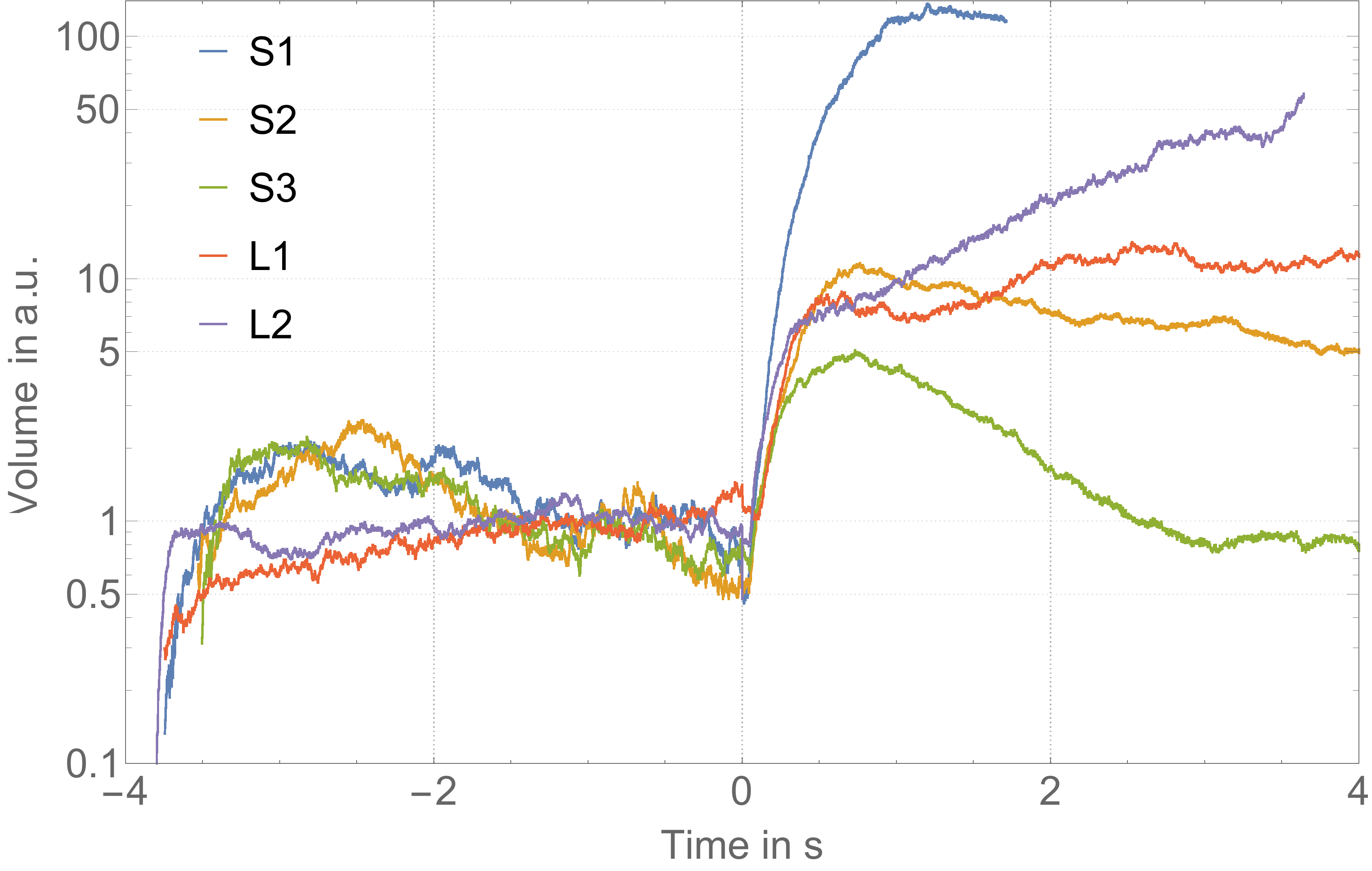}
\caption{Normalized volume of ejected particles over time. The laser is turned off at $T=0s$. The  volume at 1.5\,s is used as a marker for the increase in ejection rate. S is for the experiments with small, red laser beam, L is for the large IR spot.}
\label{fig:2}       % Give a unique label
\end{figure}

Time series of the ejected volumes in the 5 different experiments available from the drop tower are shown in fig. \ref{fig:2}. Time $T=0s$ corresponds to the time the laser is switched off. At $T\approx-4s$ the laser is switched on and the lid is opened resulting in a slight overshoot of the ejections. 

The two used geometries yield different volumes as the observed area and the active area were different. To compare the different datasets the volumes of each set were scaled to the mean volume in the last second before the laser was switched off.

The volume is proportional to the mass ejection rate when the dwelltime of a particle in the image stays constant during the experiment. The data then give a direct measurement of the increase in mass flux if the illumination is turned off.  
For the measurements with the small laser spot the amplification declines with time. We attribute this to the fact that the local heat transfer changes easily if a pit is created by removing dust. This is 
not visible in the experiments with larger spots where the ejection rate stays high over the total
observation time of 4s after the light source was turned off. Certainly, the ejection rate has to decrease eventually but this might occur on a timescale of rather 10 s not visible during the short time experiments \cite{Kelling2011}. 

To quantify the amplified flux in a single number we take the value reached 1.5 s after the light is turned off.
This is shown in Table \ref{tab:results}. The table includes a value from parabolic flights at 0.3 g as described below.

\begin{table}
\caption{Amplification factor of mass ejection rate if the light source is turned off. }
\begin{tabular}{lc}
Experiment &  Value \\
\hline
Small Spot 1& 136\\
Small Spot 2& 12\\
Small Spot 3& 5\\
Large Spot 1& 9\\
Large Spot 2& 15\\
Parabolic flight& 15\\
\end{tabular}
\label{tab:results}
\end{table}

Ignoring the outlier with an increase of a factor of 136 the ejection rate is enhanced by a factor of $10\pm 4$. 

\section{Modeling}
Seen in fig. \ref{fig:sketch} there is a (thin) upper layer (length $L_u$) and a (thicker) lower layer (lenght $L_l$).

The mass flow rate of the two layers have to be equal $\dot{M}_l=\dot{M}_u$.
Assuming that the magnitude of the temperature difference in both regions can be described by $\Delta T_u=\chi \Delta T_l$ ($T_{avg}$, $p_{avg}$ and $L_r$ taken as constant) the pressure
support on the top layer is:
\begin{equation}
\Delta p= \frac{Q_T}{Q_P}\frac{p_{avg}}{T_{avg}}\frac{L_u+\chi L_l}{L_u+L_l}\Delta T_l\,.
\end{equation}

This pressure difference can be compared for illuminated and shadowed settings. 
Temperature profiles in a dust bed under similar conditions were calculated by \cite{Kocifaj2011}. For an illuminated dust bed $\chi\approx0$. They give $\chi\approx0.06$ and $L_l/L_u=1/15$ if shadowed for 0.05\,s (the timing needed for the first ejecta in the dark) 
This results in a factor of
\begin{equation}
\frac{\Delta p_{dark}}{\Delta p_{illum}}=1+\chi \frac{L_l}{L_u}.
\end{equation}

which gives an enhancement factor of 10. Assuming that the pressure increase is correlated to the
dust mass ejection rate, this matches the data to rough order of magnitude.

\section{Scaling}

\begin{figure}
\includegraphics[width=0.5\textwidth]{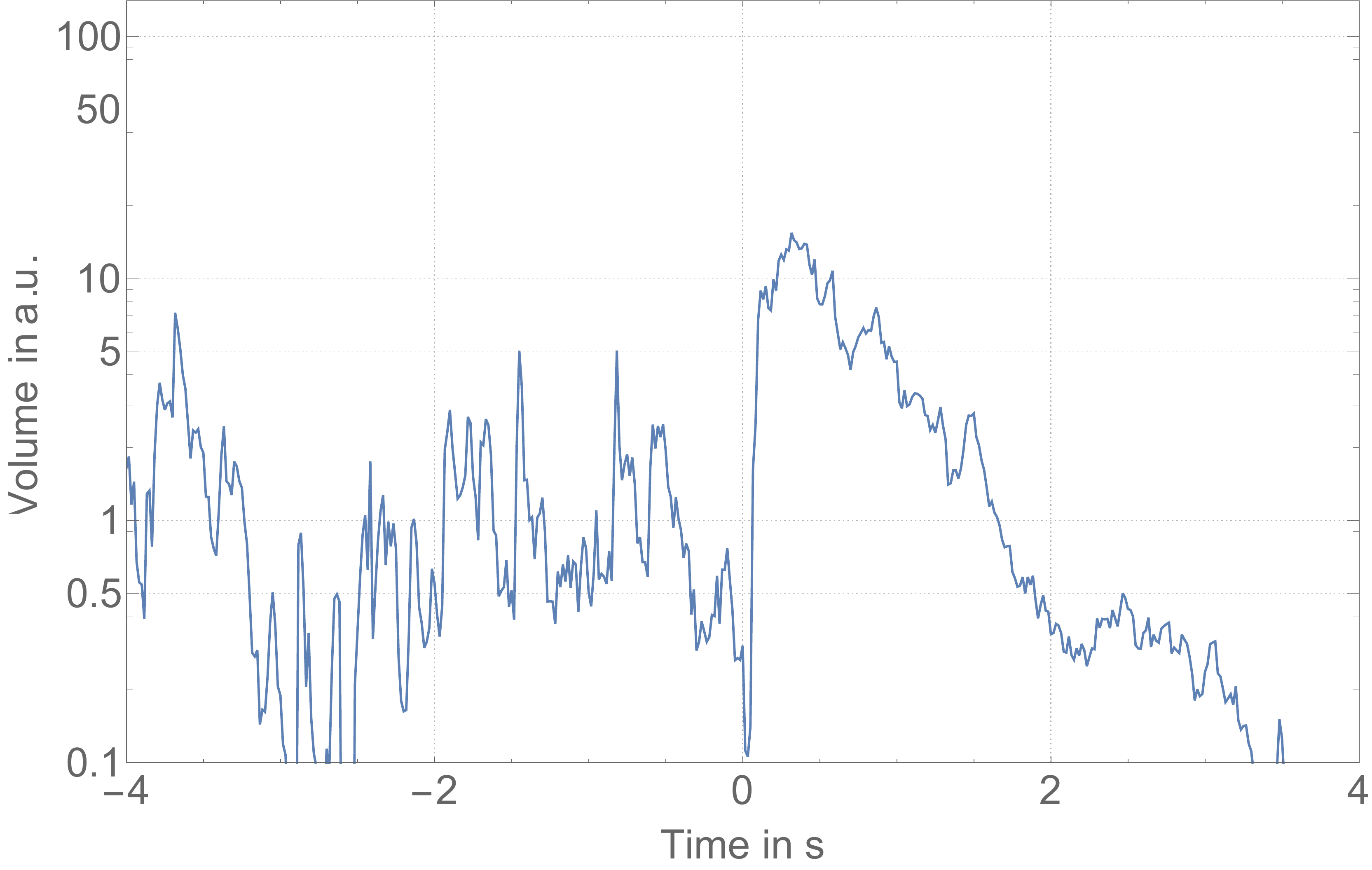}
\caption{Data from the parabolic flight experiment. Due to g-jitter and the residual gravity set in general (0.3g) the dwelltime of the particles in the field of view is much shorter and the ejections appear more spike like.}
\label{pfc}
\end{figure}

No experimental data is available yet to allow an unambiguous scaling of the amplification factor to other fluxes, dust samples or illumination spectra. Some data could be gained on the gravity dependence on parabolic flights. The setup was similar but placed on a centrifuge. We used a laser of 655 nm, a spot size of 10\,mm diameter, a light flux of 9.8\,kW/m$^2$ and worked at an ambient pressure of 4\,mbar. Here, we find an amplification factor of 15 for a similar kind of dust sample (sieved fraction between 0-125 mu m was used) at 0.3 g (see fig. \ref{pfc}). This matches the other values at 0g very well but it has to be noted that this was only one experiment that could be analyzed along the same line here. 
However, assuming the amplification is constant at different flux and gravity, an extrapolation to Martian conditions is possible. 

A free parameter is the optical thickness $\tau$ of the dust devil casting the shadow. To apply our
results to dust devils it has to be considered that most of the lifting initially is not done by the illumination but we assume it to be dominated by gas drag. 
The force $F_L(u_t)$ of the threshold wind velocity $u_t$ to lift a particle can be written as force balance with the other acting forces
\begin{equation}
F_L(u_t)=F_{Thres}=F_{g}+F_{ad}-F_{il}-F_{dk},
\end{equation}
where $F_g$ is the gravitational force on a grain, $F_ad$ the adhesion force which have to be balanced by a lifting force $F_L \propto u_t^2$, the illumination induced force $F_{il} \propto e^{-\tau}$ and the force due to the shadowing $F_{dk} \propto  (1-e^{-\tau})$. The last two were considered above.

In a wind dominated entrainment the transport rate $\dot{Q}$ scales with  
\begin{equation}
\dot{Q}\propto (u^2-u_t^2)u,
\end{equation}
where the wind speed is $u$ and $u_t$ is the threshold velocity for particle entrainment. 

As \cite{Kuepper2015} show the reduction of the threshold velocity $u_t$ due to Martian insolation is about 10\,\%~--- as  $F_L\propto u_t^2$ the reduction regarding the force is 19\,\%. By putting this in the threshold equation $F_L(u_{t,red})=F_{g}+F_{ad}-F_{il}=0.81(F_g+F_{ad})$ the relative strength of $F_{il}$ can be determined. Taking into account the enhancement by a factor 10 for shadowing the ratio between the thresholds then becomes
\begin{equation}
b=\frac{u_{t,red}}{u_t}=\sqrt{1-0.19e^{-\tau}-1.9(1-e^{-\tau})}
\end{equation}
This gives the interesting result that particle entrainment could be possible by strong shadowing ($\tau>0.64$) even without wind. An optical thick dust devil might therefore be self-sustained even if the gas drag related forces would fall below the threshold.
While the occurrence of a total shadow is unlikely, reduced flux is possible due to dust devils. 

Combined with the transport rate this yields a dust mass flow enhancement $\delta$ of a dust devil due to self shadowing. It can be expressed by the transportrate with reduced threshold velocity compared to the transport rate obtained with neglecting this reduction
\begin{equation}
\delta=\frac{\dot{Q}(u,u_{t,red})}{\dot{Q}(u,u_{t})}=\frac{a^2-b^2}{a^2-1}\,,
\end{equation}
with the factor $a=u/u_t$.
A dust devil with opacity of 0.1 is plausible \citep{Greeley2006b}, and for a wind 5\% over the threshold ($a=1.05$) this yields an amplification on Mars by a factor 4.4. We note that this enhancement strongly depends on the factor $a$, and was made for a wind dominated flux which may not always be the case.  Nevertheless, it is a plausible way to explain the difference in measured and calculated mass flows due to dust devils:
Mass fluxes estimated from observations for Gusev Crater are 19\,kg/(m$^2\,$s) \citep{Greeley2006b}. Estimations deduced from laboratory measurements by \cite{Neakrase2010} however reach only 3.66\,kg/(m$^2\,$s), an enhancement of the mass flow by a factor of 5 is therefore needed to reconcile theory and measurement. While both, experiment and observations, need assumptions to derive the mass flux some uncertainty arises. Assumptions are also needed to scale this effect to the Martian surface as the experiment was not conducted under Martian conditions, therefore some uncertainty on the absolute strength of this effect exists.

\section{Conclusions and caveats}

These first measurements are far from being a complete explanation for enhanced dust mass flow in Martian dust devils. They are but a first  set of microgravity experiments. Experiments at higher g-levels are difficult as ejected particles can fall back and escape detection. Ground experiments in wind tunnels might be feasible and are envisioned for future studies to complement pure insolation. However, the experiments stand as they are and clearly point in a direction where a change of insolation on timescales of seconds can have significant influence on soil / atmosphere interaction. 
 
This mechanism cannot explain the initial formation of a dust devil. However if it comes into existence the survival time (as 
visualized by dust) and loft rates could both be increased. The exact scaling bears more than some uncertainty and should only be regarded as order of magnitude estimate. But if we scale the results in
this simplified matter we might carefully speculate that this effect might explain the observed difference between dust loading of Martian dust devils estimated from observations and simulated.

\section{Acknowledgments}
This project is supported by DLR Space Management with funds provided by the Federal Ministry of Economics and Technology (BMWi) under grant number DLR 50 WM 1242 and DLR 50 WM 1542. Markus Kuepper is supported by the DFG. We appreciate that DLR provided access to parabolic flights and the drop tower. Our sincere thanks also go to C. de Beule for providing the data gathered during
the drop tower campaign and also to T. Jankowski who provided Figure \ref{sketchi}.


\bibliographystyle{elsarticle-harv}
\bibliography{amplification.bib}   % name your BibTeX data base

\begin{thebibliography}{14}
\expandafter\ifx\csname natexlab\endcsname\relax\def\natexlab#1{#1}\fi
\expandafter\ifx\csname url\endcsname\relax
  \def\url#1{\texttt{#1}}\fi
\expandafter\ifx\csname urlprefix\endcsname\relax\def\urlprefix{URL }\fi

\bibitem[{{Balme} and {Greeley}(2006)}]{Balme2006}
{Balme}, M., {Greeley}, R., Sep. 2006. {Dust devils on Earth and Mars}. Reviews
  of Geophysics 44, 3003.

\bibitem[{de~Beule et~al.(2014)de~Beule, {Wurm}, {Kelling}, {K{\"u}pper},
  {Jankowski}, and {Teiser}}]{deBeule2014}
de~Beule, C., {Wurm}, G., {Kelling}, T., {K{\"u}pper}, M., {Jankowski}, T.,
  {Teiser}, J., 1 2014. {The martian soil as a planetary gas pump}. Nature
  Physics 10, 17--20.

\bibitem[{{Greeley} et~al.(2006{\natexlab{a}}){Greeley}, {Arvidson}, {Barlett},
  {Blaney}, {Cabrol}, {Christensen}, {Fergason}, {Golombek}, {Landis},
  {Lemmon}, {McLennan}, {Maki}, {Michaels}, {Moersch}, {Neakrase}, {Rafkin},
  {Richter}, {Squyres}, {de Souza}, {Sullivan}, {Thompson}, and
  {Whelley}}]{Greeley2006b}
{Greeley}, R., {Arvidson}, R.~E., {Barlett}, P.~W., {Blaney}, D., {Cabrol},
  N.~A., {Christensen}, P.~R., {Fergason}, R.~L., {Golombek}, M.~P., {Landis},
  G.~A., {Lemmon}, M.~T., {McLennan}, S.~M., {Maki}, J.~N., {Michaels}, T.,
  {Moersch}, J.~E., {Neakrase}, L.~D.~V., {Rafkin}, S.~C.~R., {Richter}, L.,
  {Squyres}, S.~W., {de Souza}, P.~A., {Sullivan}, R.~J., {Thompson}, S.~D.,
  {Whelley}, P.~L., Jan. 2006{\natexlab{a}}. {Gusev crater: Wind-related
  features and processes observed by the Mars Exploration Rover Spirit}.
  Journal of Geophysical Research (Planets) 111, 2.

\bibitem[{{Greeley} et~al.(1980){Greeley}, {Leach}, {White}, {Iversen}, and
  {Pollack}}]{Greeley1980}
{Greeley}, R., {Leach}, R., {White}, B., {Iversen}, J., {Pollack}, J.~B., Feb.
  1980. {Threshold windspeeds for sand on Mars - Wind tunnel simulations}.
  Geophysical Research Letters 7, 121--124.

\bibitem[{{Greeley} et~al.(2006{\natexlab{b}}){Greeley}, {Whelley}, {Arvidson},
  {Cabrol}, {Foley}, {Franklin}, {Geissler}, {Golombek}, {Kuzmin}, {Landis},
  {Lemmon}, {Neakrase}, {Squyres}, and {Thompson}}]{Greeley2006a}
{Greeley}, R., {Whelley}, P.~L., {Arvidson}, R.~E., {Cabrol}, N.~A., {Foley},
  D.~J., {Franklin}, B.~J., {Geissler}, P.~G., {Golombek}, M.~P., {Kuzmin},
  R.~O., {Landis}, G.~A., {Lemmon}, M.~T., {Neakrase}, L.~D.~V., {Squyres},
  S.~W., {Thompson}, S.~D., Dec. 2006{\natexlab{b}}. {Active dust devils in
  Gusev crater, Mars: Observations from the Mars Exploration Rover Spirit}.
  Journal of Geophysical Research (Planets) 111~(E10), 12.

\bibitem[{{Kelling} et~al.(2011){Kelling}, {Wurm}, {Kocifaj}, {Kla{\v c}ka},
  and {Reiss}}]{Kelling2011}
{Kelling}, T., {Wurm}, G., {Kocifaj}, M., {Kla{\v c}ka}, J., {Reiss}, D., 2011.
  {Dust ejection from planetary bodies by temperature gradients: Laboratory
  experiments}. Icarus 212, 935--940.

\bibitem[{{Knudsen}(1909)}]{Knudsen1909}
{Knudsen}, M., 1909. {Eine Revision der Gleichgewichtsbedingung der Gase.
  Thermische Molekularstr{\"o}mung}. Annalen der Physik 336, 205--229.

\bibitem[{{Kocifaj} et~al.(2011){Kocifaj}, {Kla{\v c}ka}, {Kelling}, and
  {Wurm}}]{Kocifaj2011}
{Kocifaj}, M., {Kla{\v c}ka}, J., {Kelling}, T., {Wurm}, G., Jan. 2011.
  {Radiative cooling within illuminated layers of dust on (pre)-planetary
  surfaces and its effect on dust ejection}. Icarus 211, 832--838.

\bibitem[{{K{\"u}pper} and {Wurm}(2015)}]{Kuepper2015}
{K{\"u}pper}, M., {Wurm}, G., Jul. 2015. {Thermal creep-assisted dust lifting
  on Mars: Wind tunnel experiments for the entrainment threshold velocity}.
  Journal of Geophysical Research (Planets) 120, 1346--1356.

\bibitem[{{Muntz} et~al.(2002){Muntz}, {Sone}, {Aoki}, {Vargo}, and
  {Young}}]{Muntz2002}
{Muntz}, E.~P., {Sone}, Y., {Aoki}, K., {Vargo}, S., {Young}, M., 1 2002.
  {Performance analysis and optimization considerations for a Knudsen
  compressor in transitional flow}. Journal of Vacuum Science Technology A:
  Vacuum, Surfaces, and Films 20~(1), 214--224.

\bibitem[{Neakrase et~al.(submitted)Neakrase, Balme, Esposito, Kelling, Klose,
  Kok, Marticorena, Merrison, Patel, and Wurm}]{Neakrase2016}
Neakrase, L., Balme, M., Esposito, F., Kelling, T., Klose, M., Kok, J.,
  Marticorena, B., Merrison, J., Patel, M., Wurm, G., Mar. submitted. Particle
  lifting processes in dust devils. Space Science Reviews.

\bibitem[{{Neakrase} and {Greeley}(2010)}]{Neakrase2010}
{Neakrase}, L., {Greeley}, R., Mar. 2010. {Dust devil sediment flux on Earth
  and Mars: Laboratory simulations}. Icarus 206, 306--318.

\bibitem[{Reiss et~al.(2014)Reiss, Hoekzema, and Stenzel}]{Reiss2014b}
Reiss, D., Hoekzema, N., Stenzel, O., 2014. Dust deflation by dust devils on
  mars derived from optical depth measurements using th shadow method in hirise
  images. Planetary and Space Science 93, 54 -- 64.

\bibitem[{{Sone} and {Itakura}(1990)}]{Sone1990}
{Sone}, Y., {Itakura}, E., 1990. {Analysis of Poiseuille and thermal
  transpiration flows for arbitrary Knudsen numbers by a modified Knudsen
  number expansion method and their database.} Journal of the Vacuum Society of
  Japan 33, 92--94.

\end{thebibliography}

%\printbibliography

\end{document}